\documentclass [superscriptaddress,amsmath,prb,twocolumn,amssymb]{revtex4-1}
\usepackage{graphicx}
\usepackage{dcolumn}
\usepackage[none]{hyphenat} 
\usepackage{braket}
\usepackage{multirow}
\usepackage{booktabs}
\usepackage{bm,color,upgreek}
\usepackage{braket}

\begin{document}

\title{Electrically tuneable nonlinear anomalous Hall effect in two-dimensional transition-metal dichalcogenides WTe$_2$ and MoTe$_2$}

\author{Yang Zhang}
\affiliation{Max Planck Institute for Chemical Physics of Solids, 01187 Dresden, Germany}
\affiliation{Leibniz Institute for Solid State and Materials Research, IFW Dresden, 01069 Dresden, Germany}
\author{Jeroen van den Brink}
\affiliation{Leibniz Institute for Solid State and Materials Research, IFW Dresden, 01069 Dresden, Germany}
\affiliation{Institut für Theoretische Physik, TU Dresden, 01062 Dresden, Germany}

\author{Claudia Felser}
\affiliation{Max Planck Institute for Chemical Physics of Solids, 01187 Dresden, Germany}
\author{Binghai Yan}
\email{binghai.yan@weizmann.ac.il}
\affiliation{Department of Condensed Matter Physics, Weizmann Institute of Science, Rehovot, 7610001, Israel}

\begin{abstract}
We studied the nonlinear electric response in WTe$_2$ and MoTe$_2$ monolayers.
When the inversion symmetry is breaking but the the time-reversal symmetry is
preserved, a second-order Hall effect called the nonlinear anomalous Hall effect (NLAHE)
emerges owing to the nonzero Berry curvature on the nonequilibrium Fermi surface.
We reveal a strong NLAHE with a Hall-voltage that is quadratic with respect to the longitudinal current.
The optimal current direction is normal to the mirror plane in these
  two-dimensional (2D) materials.
The NLAHE can be sensitively tuned by an out-of-plane electric field, which
  induces a transition from a topological insulator to a normal insulator.
Crossing the critical transition point, the magnitude of the NLAHE increases,
  and its sign is reversed.
Our work paves the way to discover exotic nonlinear phenomena in inversion-symmetry-breaking 2D materials.
\end{abstract}

\maketitle

\section{Introduction}
The past decade has seen intensive inverstigation of band-structure topology and
the discovery of novel topological states and topological materials, such as topological insulators (TIs)~\cite{Qi2011RMP,Hasan2010RMP,Yan2012rpp,Ando2013} and topological Dirac and Weyl semimetals~\cite{Wan2011,volovik2003universe,Murakami2007,Burkov2011,Hosur2013,Yan2017,Armitage2017}.
The band-structure topology and electronic response functions are commonly
characterised by the Berry curvature~\cite{Xiao2010} in the momentum($k$) space.
A well-established example is the anomalous Hall effect~\cite{Nagaosa2010}  as
an intrinsic property originating in the band structure, in which the Berry
curvature integrated over the occupied states at the thermodynamic equilibrium
gives rise to the Hall conductivity. The anomalous Hall effect and its quantised
version have recently been observed in time-reversal symmetry (TRS) breaking
topological materials, for example, magnetically doped
TIs~\cite{Liu2008,Yu2010,Chang2013} and magnetic Weyl
semimetals~\cite{Xu2011,Yang2011QHE,Burkov2014,Yang2017}. In recent years there
has been increasing interest in the nonlinear optical and electrical properties
of topological materials, considering the Berry phase effect~\cite{Moore2010,Deyo2009,Young2012,Sipe2000,Morimoto2016a}.

In the linear response regime, the anomalous Hall effect vanishes in the
presence of TRS, because TRS forces the Berry curvature to be odd with respect
to $k$, i.e. $\Omega^n(k)=-\Omega^n(-k)$ where $n$ is the band index. In the
nonlinear response regime, however, an intriguing nonlinear anomalous Hall
effect (NLAHE) can still survive in the presence of TRS but the absence of the
inversion symmetry~\cite{Sodemann2015}. When an electric field \textbf{$E$}
drives a current through a crystal in the steady state, the system is out of
equilibrium and the Fermi surface exhibits an effective shift in $k$-space.
Therefore, the Fermi occupations at $k$ and $-k$ are no longer necessarily the
same any more. This leads to a net Berry curvature summed on the nonequilibrium
Fermi surface, i.e, an anomalous Hall conductivity that is proportional to \textbf{$E$}  and the relaxation time $\tau$. Thus, the Hall voltage is estimated to be quadratic, rather than linear, to the longitudinal electric field \textbf{$E$}. The NLAHE was derived at the zero-frequency limit of the nonlinear photocurrent generation~\cite{Sodemann2015,Morimoto2016}.
Although it is a nonequilibrium property, the NLAHE can be described by a
geometric quantity at the equilibrium Fermi surface, the Berry curvature dipole
(BCD)~\cite{Sodemann2015}. Very recently the BCD induced NLAHE was calculated
for the three-dimensional (3D) Weyl semimetals based on $ab~initio$ band
structures~\cite{Zhang2018} and also for tellurium~\cite{Tsirkin2018}.

The transition-metal dichalcogenides WTe$_2$ and MoTe$_2$ are WSMs in the 3D
bulk~\cite{Soluyanov2015WTe2,Sun2015MoTe2} and become two-dimensional (2D) TIs
in the monolayer (ML) form~\cite{Qian2014,Tang2017,Peng2017,Wu2018}. Nonlinear
optical phenomena, for example, a nonlinear photocurrent, were recently reported in their bulk systems~\cite{Lim2018,Ji2018}.
In this work, we investigate the NLAHE in MLs of WTe$_2$ and MoTe$_2$,
because the use of MLs makes it easier to tune the band structure by an external electric field, for instance, by applying a back gate~\cite{Wu2018}.
The out-of-plane electric field is found to sensitively tune the band structure
and consequently control the BCD,
which is an intrinsic property determined by the band structure and wave functions.
Near the electric-field-induced topological transition from a TI to a normal insulator,
the BCD and NLAHE are found to be strongly enhanced.

\begin{figure}[htbp]
\begin{center}
\includegraphics[width=0.5\textwidth]{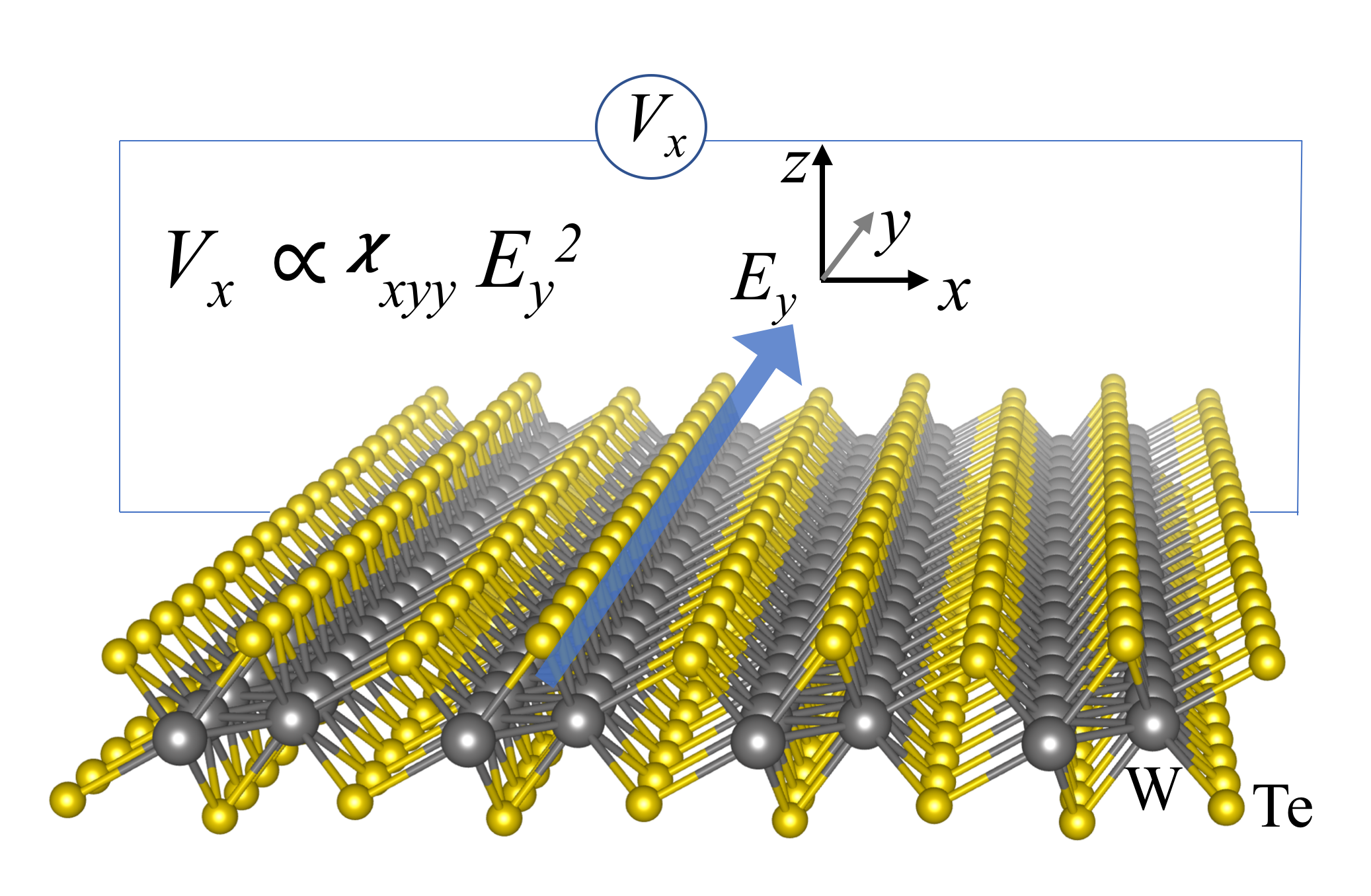}
\end{center}
\caption{
Crystal structure and the nonlinear anomalous Hall effect in the WTe$_2$ monolayer.
W and Te atoms are represented by gray and yellow spheres.
When applying an electric field along the $y$ axis, i.e. the W chain direction that crosses the mirror plane,
a Hall-like voltage $V_x$ appears and is quadratic to the electric field $E_y$.
Such a nonlinear anomalous Hall effect exists without breaking the time-reversal symmetry.
}
\label{model}
\end{figure}

\section{Methods}
The WSM state of WTe$_2$ or MoTe$_2$ refers to the $T_d$ phase of the
crystal structure (space group $Pmn2_1$, No. 31), in which inversion symmetry is
broken. However, the corresponding ML recovers the inversion symmetry by a slight distortion,
commonly called the $1T^{\prime}$ phase (space group $P2_1/m$, No. 11). This inversion
symmetry can be broken by applying an out-of-plane electric field ($E$) or even by
the existence of a substrate. We focus on the $1T^{\prime}$-MLs of the two
compounds under different electric fields for two compounds. In addition,
we also investigate the $T_d$-MLs for comparison, considering the fact that
the phase transition may occur under special conditions (e.g.
Refs~\onlinecite{Qi2016MoTe2,Yang2017elastic}). Both $T_d$- and
$1T^{\prime}$-MLs share a mirror plane $\mathcal{M}_y: y \rightarrow - y$. A
zigzag-shape Mo or W atomic chain forms and crosses the mirror plane. The
$\mathcal{M}_y$ symmetry is crucial to determining the symmetry of the NLAHE as we will discuss.

We first perform $ab~initio$ density-functional theory (DFT) calculations and
then project the DFT band structure to atomic-like Wannier functions with FPLO~\cite{koepernik1999}. Starting with the one-particle Hamiltonian ($\hat{H}$) in the basis of Wannier functions,
we compute the distribution of the Berry curvature $\Omega^n(\mathbf{k})$ in the
momentum space ($\mathbf{k}$). In a 2D system, $\Omega^n(\mathbf{k})$ only has a $z$ component~\cite{Xiao2010},
 \begin{equation}
\label{eq:berry}
\begin{aligned}
\Omega^n_z(\mathbf{k})= 2i \hbar^2 \sum_{m \ne n} \frac{\bra{n} \hat{v}_x \ket{m} \bra{m} \hat{v}_y \ket{n}}{(\epsilon_{n}-\epsilon_{m})^2},
\end{aligned}
\end{equation}
where $\epsilon_{n}$ and $\ket{n}$ are eigenvalues and eigen wave functions, respectively, of $\hat{H}$ at the momentum $\mathbf{k}$ and $\hat{v}_{x,y}=\frac{d\hat{H}} {\hbar dk_{x,y}}$ the velocity perator.

The nonlinear responses includes the dc current $\textit{j}_a^{(0)}=\chi_{abc} \mathcal{E}_b \mathcal{E}_c^*$ and the second harmonic generation $\textit{j}_a^{(2\omega)}=\chi_{abc} \mathcal{E}_b \mathcal{E}_c$, under the oscillating electric field $E_c(t)=Re\{ \mathcal{E}_c e^{i \omega t}\}$ of light, where $a,b,c = x,y,z$.
At the zero-$\omega$ limit, the dc current still preserves $\textit{j}_a= 2
\textit{j}_a^{(0)} |_{\omega \rightarrow 0} = 2 \chi_{abb} |\mathcal{E}_b|^2$,
leading to the so-called NLAHE~\cite{Sodemann2015}.
Although the NLAHE is related to the net Berry curvature due to a nonequilibrium
Fermi distribution,  the nonlinear conductivity $\chi_{abb}$ can be describe by
the BCD, a quantity defined in the equilibrium state in the semiclassical approximation~\cite{Sodemann2015} as follows,
\begin{eqnarray}
\label{eq:chi}
\chi_{abb} &=& - \varepsilon_{adb} \frac{e^3\uptau}{2 \hbar^2 (1+ i \omega \uptau)} D_{bd}\\
\label{eq:dipole}
D_{bd} &=& \int_k f^0_n(\mathbf{k}) \frac{\partial{\Omega^n_d}}{\partial{k_b}},
\end{eqnarray}
where $D_{bd}$ is the BCD, $f^0_n(\mathbf{k})$ the equilibrium Fermi--Dirac
distribution, $\uptau$ the relaxation time, $\varepsilon_{adb}$  the third rank
Levi--Civita symbol, $a,b=x,y$ and $d=z$ in 2D. We compute $\Omega_d$ by
Eq.~\ref{eq:berry} and then calculate the $D_{bd}$ by integrating
$\frac{\partial{\Omega^n_d}}{\partial{k_b}}$ in a very dense $k$-grid ($2000
\times 2000$) to obtain converged values of the BCD. The BCD is dimensionless in
three dimensions, whereas it is in unit of length in 2D.

Because $\Omega^n_z$ is odd with respect to the $\mathcal{M}_y$ reflection,
$\frac{\partial{\Omega^n_z}}{\partial{k_x}}$ is odd to $\mathcal{M}_y$ while
only $\frac{\partial{\Omega^n_z}}{\partial{k_y}}$ is even. Therefore, only
$D_{yz}$ and $\chi_{xyy}$ are nonzero. Thus, the nonlinear Hall voltage $V_x$
appears inside the mirror plane when an electric field $E_y$ passes along the W
or Mo chains, and $V_x \propto \chi_{xyy} E_y^2$, as illustrated in Fig. 1. If
the voltage contacts and electric field directions are switched, there will be no
NLAHE signal. This a strong anisotropy can serve as a useful tool to distinguish the NLAHE from other effects in WTe$_2$ and MoTe$_2$ MLs.

\section{Results and Discussion}

\subsection{ $1T^{\prime}$ Monolayers of WTe$_2$}

We start with the $1T^{\prime}$-ML of WTe$_2$. It is known to be a 2D TI. An
applied electric field breaks the inversion symmetry and induce a transition
from a TI to a trivial insulator. During the transition, the band gap first shrinks to
zero and then opens again.
A previous study on the TaAs-family of WSMs~\cite{Zhang2018} revealed that the small gap
or zero gap region contributes a large gradient of the Berry curvature, i.e, a large BCD.
This can be intuitively understood from Eq.~\ref{eq:berry}. The smallness of the
energy gap (i.e, the denominator of Eq.~\ref{eq:berry}) usually indicates that a
large Berry curvature is concentrated on the small gap region, which is usually a
narrow momentum area, leading to a large gradient of the Berry curvature.
In addition to a large magnitude of the BCD, a sign change of the BCD may be induced by the phase transition, because the band inversion switches the sign of the Berry curvature.
Therefore, we are particularly interested in the evolution of the BCD with respect to the topological phase transition.

Figure 2 shows the band structures of the $1T^{\prime}$-ML under various electric
fields. At $E=0$, all the bands are doubly degenerate becomes of the coexistence of
TRS and the inversion symmetry. The band structure exhibits direct band gaps
with an inversion between the conduction and valence bands, giving rise to the 2D TI
phase. The zero indirect band gap is due to the known DFT underestimation,
whereas the quasiparticle energy correction can lift the indirect gap, as discussed in
previous work~\cite{Qian2014}. Because this gap underestimation does not affect
the Berry curvature effect investigated here, for simplicity, we use the DFT band structures in this work.
As $E$ increases, we indeed observe that the direct energy gap shrinks to zero at $E=0.0075$ V/$a_0$ ($a_0$
is the Bohr radius, 0.53 \AA) and opens again, inducing the topological transition. We note
that the band touching point at the critical electric field is not equivalent to a Weyl point,
since it is not stable against a weak perturbation (e.g. the variation of $E$) and the Weyl point
is well-defined only for 1D and 3D systems.
The Z$_2$ topological invariant before and after the transition is verified by tracing the
Wannier centres by the Wilson loop method~\cite{Yu2011,Soluyanov2011}.

\begin{figure}[htbp]
\begin{center}
\includegraphics[width=0.5\textwidth]{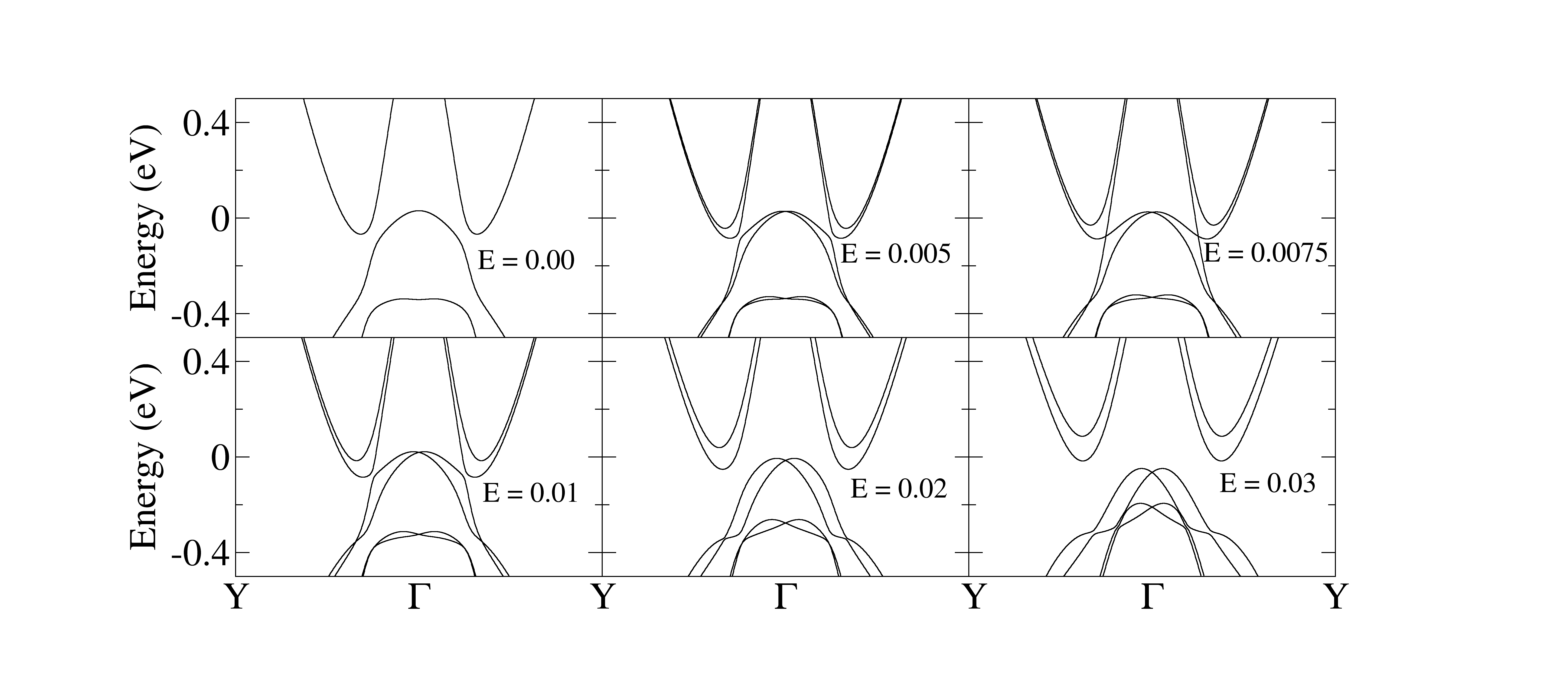}
\end{center}
\caption{
Band structures of WTe$_2$ $1T^{\prime}$-monolayers with applied electric field. The electric field $E$ is out-of-plane and in unit of eV/$a_0$,
where $a_0$ is the Bohr radius.
The band dispersion is shown along the $\Gamma$--Y direction, i.e. along the zigzag W chain.
}
\label{model}
\end{figure}

We show the corresponding BCD in Figure 3. At $E=0$, the BCD vanishes regardless
of position of
the Fermi energy ($E_F$),
because the coexistence of TRS and inversion symmetry forces $\Omega(\mathbf{k}) = 0$ at every momentum $\mathbf{k}$.
At $E=0.005$ V/$a_0$ (the TI phase), a large BCD appears (Fig. 3a).
Because it is a Fermi surface property, the BCD depends sensitively on the position of $E_F$.
We focus on the BCD for $E_F = 0$, i.e.  the charge neutral point.
As $E$ increases from zero, the BCD varies nonmonotonically and is characterised by four regimes.
It (i) first increases in a positive amplitude (e.g. $E=0.005)$, (ii) then reduces to zero (near $E=0.0075$),  (iii) further grows with a negative amplitude (e.g. $E=0.01$), and (iv) finally decreases to zero at large $E$($E>0.03$).
The regime-(i) is caused by the emergence of nonzero BCD by breaking the inversion symmetry
with $E$. Because the phase transition switches the order of the conduction and valence bands,
the sign of the Berry curvature is reversed by the transition, and thus the sign
of the BCD is also reversed for regimes-(ii) and (iii).
This behaviour is quite similar to the sign change of the photocurrent calculated at the topological phase transition of TIs~\cite{tan2016enhancement}.
At large $E$, the system becomes a gapped insulator; thus, the Fermi surface
vanishes at $E_F=0$ and the BCS then becames zero.
This behaviour explains the regime-(iv). For $E=0.005$ and $0.01$, one can find a peak of
BCD slightly below $E_F=0$ in Fig. 3. This is because the smallest energy-gap (
the largest Berry curvature) appears slightly below $E_F=0$.

When the BCD is projected to 2D $k$-space, it is easier to understand the BCD
from the corresponding band structure.
For example, at $E=0.005$ the BCD is predominantly contributed by the small-gap
region, for example, the nearly band touching positions along the $\Gamma$-Y line.
After the band inversion at $E=0.01$, the BCD indeed changes sign.
As $E$ increases further to $E=0.02$ and $E=0.03$, the BCD decreases exponentially in amplitude, corresponding to regime-(iv) discussed above.

\begin{figure}[htbp]
\begin{center}
\includegraphics[width=0.5\textwidth]{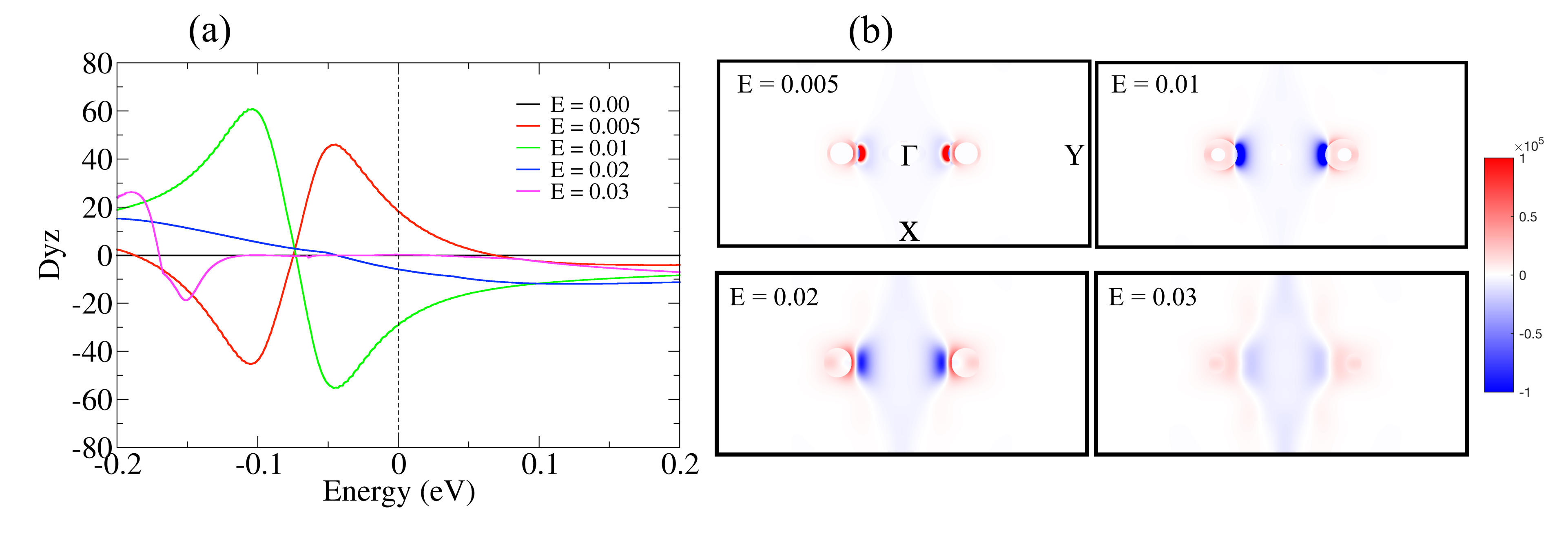}
\end{center}
\caption{
The Berry curvature dipole at different Fermi energy and its distribution in the 2D Brillouin zone for WTe$_2$ $1T^{\prime}$-monolayers with varying applied electric field.
The Berry curvature dipole is in unit of $a_0$, where $a_0$ is the Bohr radius 0.529 \AA.
In the 2D Brillouin zone, red and blue colors present positive and negative
  amplitudes (in arbitrary units) of the Berry curvature dipole.
}
\label{model}
\end{figure}

\subsection{$T_d$ Monolayers of WTe$_2$}

The $T_d$-ML breaks inversion symmetry. We can divide the effect of inversion symmetry breaking into
two parts, the in-plane distortion $\Delta_{\|}$ and the out-of-plane distortion $\Delta_{\bot}$.
The induced band splitting can be observed at the bottom of the conduction band (Fig. 4a). It is
also a 2D TI in topology.
When an electric field is applied along the $\Delta_{\bot}$, the symmetry breaking can be
further enhanced to be $\Delta_{\|}+E$.
For example, both the conduction and valence bands split further as $E$
increases from 0 to 0.05.  As in the $1T^\prime$ structure, $E$ drives the system to
the topological phase transition in the band structure. However, the critical field
for the $T_d$ phase ($\sim 0.10$) is much larger than that for the  $1T^\prime$ phase. Therefore, the BCD increases monotonically from $E=0$ to $E=0.05$, as shown in Fig. 4b.
When $E$ is opposite to $\Delta_{\bot}$, the effective out-of-plane inversion-symmetry-breaking is $\Delta_{\bot}-E$. As $E$ increases in amplitude, the symmetry-breaking effect first decreases and then increase again. For example, the band splitting is partially suppressed for $E=-0.01$ and is further enhanced for $E=-0.05$. The BCD shows the same trend (see Fig. 4b).

\begin{figure}[htbp]
\begin{center}
\includegraphics[width=0.5\textwidth]{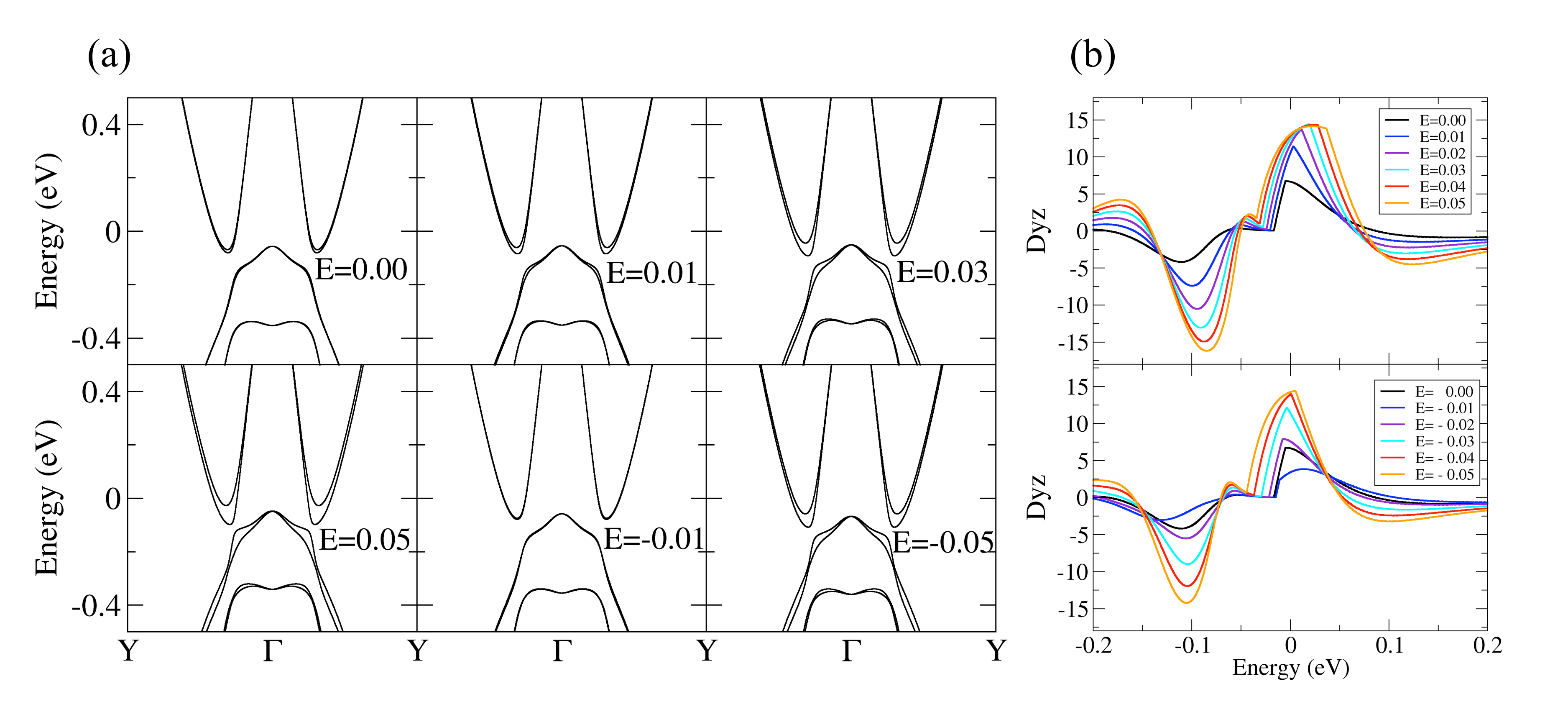}
\end{center}
\caption{Band structures and the Berry curvature dipole (in unit of $a_0$) at different electric field ($E$ in unit of V/$a_0$) for WTe$_2$ $T_d$-monolayers. }
\label{model}
\end{figure}

\subsection{Monolayers of MoTe$_2$}

The MoTe$_2$ MLs ($1T\prime$ and $T_d$) exhibit trends quite similar to those of
WTe$_2$ MLs. For the $1T\prime$ structure, the topological phase transition
occurs between $E=0.01$ and $E=0.015$. Therefore, the BCD exhibits large
magnitudes with opposite signs under these two electric fields. When $E$ increases further,
the BCD decreases in amplitude, as shown in Fig. 5.

\begin{figure}[htbp]
\begin{center}
\includegraphics[width=0.4\textwidth]{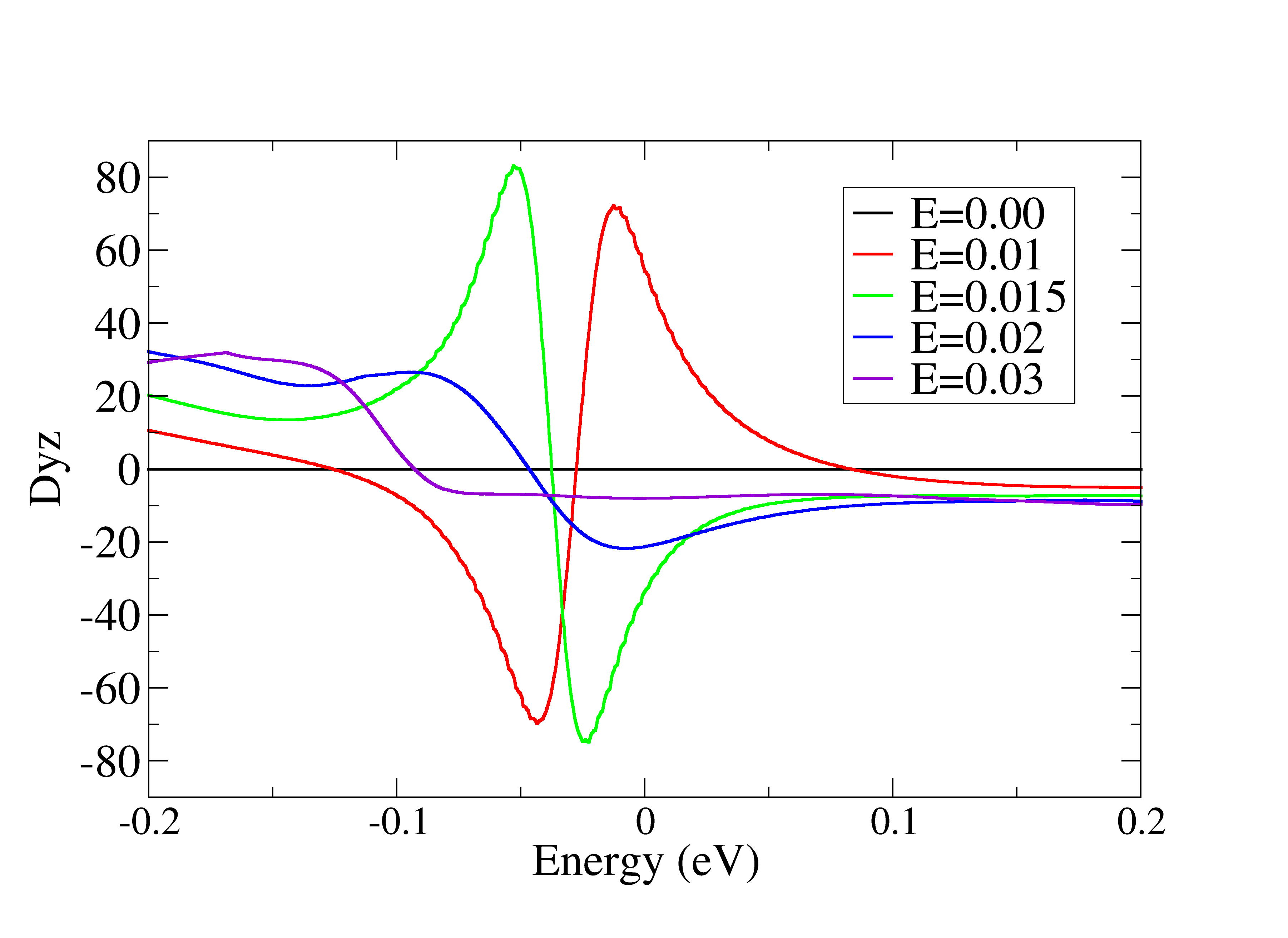}
\end{center}
\caption{
The Berry curvature dipole at different Fermi energy for MoTe$_2$ $1T^{\prime}$-monolayers with varying applied electric field.
}
\label{model}
\end{figure}

\subsection{Discussions}

We estimate the magnitude of the NALHE. In a longitudinal dc field $E_y$, the nonlinear Hall
$\textit{j}_x= 2 \textit{j}_x^{(0)} |_{\omega \rightarrow 0} = 2 \chi_{xyy} |\mathcal{E}_y|^2 = \sigma_{xy} E_y$,
where we define a Hall conductance $\sigma_{xy} \equiv 2 \chi_{xyy} E_y = \frac{e^3 \tau}{\hbar^2} D_{yz} = G_0 (\tau D_{yz} e E_y \pi/\hbar$) and $G_0 \equiv \frac{2e^2}{h}$ the conductance quantum.
Considering $E_y \sim 10^3$ V/m, $\tau \sim 1$ ps and $D_yz \sim 20 a_0$ for $1T^\prime$ WTe$_2$,
we obtain $\sigma_{xy} \sim 1\% G_0$. The magnitude of the NLAHE in MoTe$_2$-MLs
is comparable to that of WTe$_2$, as indicated in Fig. 5. This Hall conductance
can be measured under current experimental conditions.

\section{Conclusions}
In conclusion, we reveal a strong NLAHE in WTe$_2$ and MoTe$_2$ MLs.
An out-of-plane external electric field can break the inversion symmetry (for the $1T^\prime$
structure) and generate a nonzero BCD, leading to the NLAHE.
Near the topological phase transition region induced by the electric field, the NLAHE
is strongly enhanced.
In addition to a dc electric field, a longitudinal (in-plane) electric field can
also be the electric field of a low-frequency stimulation (e.g. a microwave), which may
induce an even stronger NLAHE owing to the strong electric field.
In addition to an out-of-plane electric field, the inversion symmetry of MLs can also
be broken by other ways, such as the strain (e.g. recently realised in the MoS$_2$ ML~\cite{lee2017valley}) and
the substrate proximity.
Note that our conclusions can be generalised from MLs to few layers and also to other phases (e.g. $2H$ and $1T$) of transition-metal dichalcogenides,
when the inversion symmetry as well as the three-fold rotational symmetry is broken in these systems.

\section{Acknowledgments}
Y.Z., J.vdB. and C.F. thank financial support by the German Research Foundation (DFG, SFB 1143). C.F. acknowledges the European Research Council (ERC) Advanced Grant (No. 742068) ``TOPMAT''.  B.Y. is supported by a research grant from the Benoziyo Endowment Fund for the Advancement of Science.

%

\end{document}